\documentclass[11pt,a4paper]{article}
\usepackage[T1]{fontenc}
\usepackage[utf8]{inputenc}
\usepackage{lmodern}
\usepackage[margin=25mm]{geometry}
\usepackage{amsmath,amssymb,amsthm,bm,booktabs,array}
\usepackage{cite,microtype}
\usepackage[colorlinks=true,linkcolor=blue,citecolor=blue,urlcolor=blue]{hyperref}
\usepackage{bookmark}
\hypersetup{pdftitle={Horizon Residues, Matter-Trace Winding and Test-Scalar Modes in f(Q,T) Gravity},pdfauthor={Wen-Xiang Chen}}
\allowdisplaybreaks[2]
\newcommand{\Tr}{\mathcal{T}}
\newcommand{\Res}{\operatorname{Res}}
\newcommand{\ord}{\operatorname{ord}}
\newcommand{\dd}{\mathrm{d}}
\newcommand{\ii}{\mathrm{i}}
\newcommand{\LC}{\mathring{\nabla}}
\newcommand{\eff}{\mathrm{eff}}
\newtheorem{proposition}{Proposition}

\title{\bfseries Horizon Residues, Matter-Trace Winding and Test-Scalar Modes
in \texorpdfstring{$f(Q,T)$}{f(Q,T)} Gravity}
\author{Wen-Xiang Chen$^{1}$ and M. Zeeshan Gul$^{2,3,4}$\\
\small $^{1}$School of Electronic Information, Guangzhou City University of Technology,\\$^{1}$wxchen4277@qq.com\\
\small $^2$ Department of Mathematics and Statistics, The
University of Lahore, \\
\small 1-KM Defence Road Lahore-54000, Pakistan\\
\small $^{3}$ Research Center of Astrophysics and Cosmology, Khazar
University, \\
\small Baku, AZ1096, 41 Mehseti Street, Azerbaijan\\
\small $^{4}$ Jadara Research Center, Jadara University, Irbid 21110, Jordan\\$^2$mzeeshangul.math@gmail.com}

\begin{document}
\maketitle

\begin{abstract}
We extend a local horizon-residue analysis from quadratic nonmetricity
gravity to a specified matter-trace theory,
$f(Q,\Tr)=Q+\alpha Q^2+16\pi\lambda\Tr$, where
$\Tr=g^{\mu\nu}T_{\mu\nu}$ is defined by the original matter action.
The vacuum logarithmic connection-hair branch is inherited without an
explicit $\lambda$ correction. To obtain a nonzero trace, we introduce
polynomial nonlinear electrodynamics and construct an exact magnetic
branch in coincident Cartesian Kerr--Schild coordinates. Its
nonmetricity scalar vanishes, while its matter trace is proportional to
$r^{-8}$. The metric is
$F=1-2M/r+q^2/r^2+d/r^6$, with
$d=\zeta(1-4\lambda)q^4/10$; thus the quadratic nonmetricity coupling is
inactive on this branch. We derive the metric and matter equations,
Euclidean temperature, simple-horizon residue, scalar master equation,
and first-order eikonal shifts. The logarithmic derivative
$\Tr'/\Tr$ assigns winding $-8$ to a contour enclosing the central pole,
but zero winding to a sufficiently small horizon contour. Temperature
and scalar propagation nevertheless vary continuously with $d$.
Consequently trace winding classifies divisor data rather than replacing
the matter trace or determining its backreaction. The analysis also
exhibits the degeneracy between the trace coupling and the nonlinear
electromagnetic coefficient. These are branch-specific results; no
full coupled gravitational spectrum or priority claim for the magnetic
metric is made.
\end{abstract}


\section{Problem and scope}

Symmetric teleparallel gravity uses a flat, torsion-free independent
connection and a nonmetricity scalar in place of the Levi-Civita
curvature scalar \cite{Jimenez2018,DAmbrosio2022}.
The $f(Q,T)$ extension adds dependence on the trace of the matter
energy-momentum tensor \cite{Xu2019}. Throughout this paper we write
that trace as $\Tr$ to distinguish it from the Hawking temperature $T_H$.
Its meaning is
\begin{equation}
 T_{\mu\nu}=-\frac{2}{\sqrt{-g}}
 \frac{\delta(\sqrt{-g}\mathcal L_m)}{\delta g^{\mu\nu}},
 \qquad \Tr=g^{\mu\nu}T_{\mu\nu}.
 \label{trace-definition}
\end{equation}
In particular, $\Tr$ is not the torsion scalar of $f(T)$ gravity.

For a simple static horizon, Euclidean regularity relates the Hawking
temperature \cite{Hawking1975} to a local inverse-metric residue.
The Euler-characteristic approach of Robson, Villari and Biancalana
\cite{Robson2019} and the residue approach of Ref.~\cite{Chen2025}
motivate expressing this relation in complex-analytic language.
The same pole fixes the logarithm in the scalar tortoise coordinate.
Our question is: \emph{how can matter-trace winding be included in this
description without discarding the metric and matter field equations?}

We answer by studying two distinct branches of one specified model.
The quadratic-$f(Q)$ calculation of Ref.~\cite{Chen2026} gives a vacuum
logarithmic branch. A separate, exactly solvable magnetic branch provides
$\Tr\ne0$. We do not add their metric functions: such a superposition
would require a new coupled metric--connection solution.
The precise extension is a comparison of horizon residues, the divisor
of the complexified matter trace, and scalar propagation in a branch
whose complete background equations can be checked.

There are close precedents. Magnetic solutions of Einstein gravity
coupled to nonlinear electrodynamics are well established
\cite{Bronnikov2001}; trace-dependent gravity has a corresponding
curvature formulation \cite{Harko2011}. In particular, magnetic
$r^{-6}$ corrections in trace-coupled Euler--Heisenberg models have
already been studied by Liang, Tao and Yang \cite{Liang2026}.
We therefore use this radial structure as a controlled example, rather
than claim that the metric itself is unprecedented. The focus here is
its explicit symmetric teleparallel realization and the limitations of
reconstructing thermal or wave data from trace winding.

\section{Conventions and variational equations}
\subsection{Geometry, signs and dimensions}

We use signature $(-,+,+,+)$ and $G=c=\hbar=k_B=1$.
The independent connection obeys
\begin{equation}
 R^\alpha{}_{\beta\mu\nu}(\Gamma)=0,
 \qquad \Gamma^\alpha{}_{[\mu\nu]}=0,
 \qquad Q_{\alpha\mu\nu}=\nabla_\alpha g_{\mu\nu}.
\end{equation}
Here $\nabla$ uses $\Gamma$, whereas $\LC$ and
$\mathring G_{\mu\nu}$ use the Levi-Civita connection.
Define
\begin{align}
 Q_\alpha&=g^{\mu\nu}Q_{\alpha\mu\nu},&
 \widetilde Q_\alpha&=g^{\mu\nu}Q_{\mu\alpha\nu},\\
 P^\alpha{}_{\mu\nu}
 &=-\frac14Q^\alpha{}_{\mu\nu}
 +\frac12Q_{(\mu}{}^\alpha{}_{\nu)}
 +\frac14(Q^\alpha-\widetilde Q^\alpha)g_{\mu\nu}
 -\frac14\delta^\alpha{}_{(\mu}Q_{\nu)}.
 \label{P-definition}
\end{align}
Parentheses include a factor $1/2$. We fix the convention
\begin{align}
 Q&=Q_{\alpha\mu\nu}P^{\alpha\mu\nu}\nonumber\\
 &=-\frac14Q_{\alpha\mu\nu}Q^{\alpha\mu\nu}
 +\frac12Q_{\alpha\mu\nu}Q^{\mu\alpha\nu}
 +\frac14Q_\alpha Q^\alpha
 -\frac12Q_\alpha\widetilde Q^\alpha .
 \label{Q-definition}
\end{align}
With the Einstein-tensor convention used below, the action
$+\int\sqrt{-g}Q/(16\pi)$ gives
$\mathring G_{\mu\nu}=8\pi T_{\mu\nu}$; this is the convention of
the Einstein form in Ref.~\cite{DAmbrosio2022}.
The opposite scalar convention, $Q_X=-Q$, is used in
Ref.~\cite{Xu2019}. Our model is then written
$f_X(Q_X,\Tr)=-Q_X+\alpha Q_X^2+16\pi\lambda\Tr$.
This translation is necessary: the contraction $-Q_{\alpha\mu\nu}
P^{\alpha\mu\nu}$ and the displayed expansion in
Eq.~\eqref{Q-definition} cannot be imposed simultaneously with the
same $P$.

The model and action are
\begin{equation}
 I=\int\dd^4x\sqrt{-g}
 \left[\frac{f(Q,\Tr)}{16\pi}+\mathcal L_m\right],
 \qquad f(Q,\Tr)=Q+\alpha Q^2+16\pi\lambda\Tr .
 \label{action}
\end{equation}
Flatness and vanishing torsion are imposed by constrained connection
variations, or equivalently by Lagrange multipliers.
We assume $\mathcal L_m$ and $\Tr$ have no independent-connection
dependence, as is true for the matter model below.
The dimensions are
\begin{equation}
 [Q]=[\Tr]=L^{-2},\quad [\alpha]=[\zeta]=L^2,
 \quad [\lambda]=1,\quad [M]=[q]=L,\quad [d]=L^6,
 \quad [T_H]=L^{-1}.
 \label{dimensions}
\end{equation}

\subsection{Metric variation, including the trace response}

The first two elementary variations are
\begin{equation}
 \delta\sqrt{-g}=-\frac12\sqrt{-g}g_{\mu\nu}\delta g^{\mu\nu},
 \qquad
 \delta\Tr=D_{\mu\nu}\delta g^{\mu\nu},
 \quad D_{\mu\nu}=T_{\mu\nu}+\Theta_{\mu\nu},
 \label{trace-variation}
\end{equation}
where
$\Theta_{\mu\nu}=g^{\alpha\beta}\delta T_{\alpha\beta}/\delta
g^{\mu\nu}$. Holding the connection fixed and integrating the
$\delta Q$ term by parts gives the Einstein representation of the
metric Euler derivative \cite{DAmbrosio2022}:
\begin{equation}
 E^{(Q)}_{\mu\nu}
 =f_Q\mathring G_{\mu\nu}
 +\frac12g_{\mu\nu}(Qf_Q-f)
 +2P^\alpha{}_{\mu\nu}\partial_\alpha f_Q .
 \label{EQ}
\end{equation}
The coefficient of $\delta g^{\mu\nu}$ in the full variation is
\begin{equation}
 \delta I\big|_g=\frac{1}{16\pi}\int\sqrt{-g}
 \left[E^{(Q)}_{\mu\nu}+f_{\Tr}D_{\mu\nu}
       -8\pi T_{\mu\nu}\right]\delta g^{\mu\nu}\dd^4x,
 \label{variation}
\end{equation}
up to the boundary term. This yields
\begin{equation}
 f_Q\mathring G_{\mu\nu}
 +\frac12g_{\mu\nu}(Qf_Q-f)
 +2P^\alpha{}_{\mu\nu}\partial_\alpha f_Q
 +f_{\Tr}D_{\mu\nu}=8\pi T_{\mu\nu}.
 \label{general-field}
\end{equation}
The $f_{\Tr}D_{\mu\nu}$ term is the essential trace variation
\cite{Xu2019,Harko2011}; imposing an on-shell value of $\Tr$ before
varying would in general lose this term.
For Eq.~\eqref{action},
\begin{equation}
 f_Q=1+2\alpha Q,\qquad f_{\Tr}=16\pi\lambda,
 \qquad Qf_Q-f=\alpha Q^2-16\pi\lambda\Tr.
\end{equation}
Substitution gives
\begin{equation}
 (1+2\alpha Q)\mathring G_{\mu\nu}
 +\frac{\alpha}{2}Q^2g_{\mu\nu}
 +4\alpha P^\rho{}_{\mu\nu}\partial_\rho Q
 =8\pi T^{\eff}_{\mu\nu},
 \label{model-field}
\end{equation}
with
\begin{equation}
 T^{\eff}_{\mu\nu}
 =T_{\mu\nu}+\lambda\Tr g_{\mu\nu}-2\lambda D_{\mu\nu}.
 \label{effective-tensor}
\end{equation}
An independent way to obtain the right-hand side is to vary the exact
effective matter Lagrangian
\begin{equation}
 \mathcal L_\eff=\mathcal L_m+\lambda\Tr[\mathcal L_m].
 \label{effective-action}
\end{equation}
The trace inside the brackets is always that of the \emph{original}
$\mathcal L_m$, not the trace of $\mathcal L_\eff$.

\subsection{Connection equation and the limits of vacuum inheritance}

In the absence of hypermomentum the constrained connection equation is
\begin{equation}
 \nabla_\mu\nabla_\nu
 \left[\sqrt{-g}\,f_Q P^{\mu\nu}{}_{\rho}\right]=0,
 \label{connection}
\end{equation}
where $P^{\mu\nu}{}_{\rho}=g^{\nu\sigma}P^\mu{}_{\sigma\rho}$
\cite{Xu2019,DAmbrosio2022}. To see the origin of this equation,
keep the metric fixed and use a locally allowed flat, torsion-free
connection variation:
\begin{equation}
 \delta Q=-4P^{\alpha\mu}{}_{\rho}\delta\Gamma^\rho{}_{\alpha\mu},
 \qquad
 \delta\Gamma^\rho{}_{\alpha\mu}=\nabla_\alpha\nabla_\mu\xi^\rho.
\end{equation}
Two integrations by parts in the connection part of $\delta I$ give
Eq.~\eqref{connection}. For constant $f_Q$ the equation is the
symmetric teleparallel identity of the linear-$Q$ action. In a
$Q=0$ configuration of the present model, $f_Q=1$ everywhere, so that
this identity applies even when $\Tr$ varies.

If matter is absent, then
$T_{\mu\nu}=D_{\mu\nu}=\Tr=0$ and Eqs.~\eqref{model-field} and
\eqref{connection} reduce to the quadratic-$f(Q)$ vacuum system.
Thus vacuum solutions are inherited. The statement
``$\Tr=0$'' alone is insufficient: a nonzero
$D_{\mu\nu}$ can still contribute, and matter perturbations can
contribute at second order. For Maxwell theory in four dimensions,
however, the trace vanishes as an off-shell identity and its entire
linear trace-coupling term vanishes.
These inheritance arguments do not extend unchanged to
$f=Q+\alpha Q^2+\gamma Q\Tr$: in that model
$f_Q=1+2\alpha Q+\gamma\Tr$ can have a nonzero derivative on $Q=0$.

\section{Relation to the supplied vacuum logarithmic branch}

The connection-hair solution of Ref.~\cite{DAmbrosio2022}, also used
in Ref.~\cite{Chen2026} and the supplied source manuscript, has the displayed local form
\begin{equation}
 F_v(r)=1-\frac{2M_v+\eta\ln(r/r_\star)}{r},
 \qquad \eta=\alpha^2\mu,
 \label{vacuum-F}
\end{equation}
through its stated perturbative order, after mass-parameter
renormalization. It belongs to the matter-free sector of
Eq.~\eqref{action}; its explicit dependence on $\lambda$ is therefore
zero. The omitted higher-$\alpha$ terms are not fixed by
Eq.~\eqref{vacuum-F}.
At a simple root $r_v$,
\begin{align}
 r_v&=2M_v+\eta\ln(r_v/r_\star),\\
 F_v'(r)&=\frac{2M_v+\eta[\ln(r/r_\star)-1]}{r^2},\\
 F_v'(r_v)&=\frac{r_v-\eta}{r_v^2},\qquad
 \Res_{r_v}\frac{1}{F_v}=\frac{r_v^2}{r_v-\eta},\\
 T_{H,v}&=\frac{r_v-\eta}{4\pi r_v^2}.
 \label{vacuum-result}
\end{align}
These last expressions are algebraic identities for the displayed
function, with their physical use restricted to the controlled
perturbative region and the chosen normalization of $t$.
Writing $L_v=\ln(2M_v/r_\star)$ and substituting
$r_v=2M_v+\eta a+\cdots$ into the first equation gives
$a=L_v$, and hence
\begin{equation}
 r_v=2M_v+\eta L_v+\cdots,\qquad
 T_{H,v}=\frac{1}{8\pi M_v}
 \left[1-\frac{\eta}{2M_v}(L_v+1)\right]+\cdots .
 \label{vacuum-expansion}
\end{equation}
The dots include uncomputed orders of the original coupling expansion,
not only powers $\eta^2$ generated by formally resumming the metric.
We require small corrections to both the metric and its derivatives
near the horizon; the far-zone condition
$|\eta\ln(r/r_\star)|\ll 2M_v$ is also needed to keep the displayed
tail small relative to the Schwarzschild tail.

The function $r[1-F_v(r)]/2=M_v+\eta\ln(r/r_\star)/2$ has no finite
large-$r$ limit for nonzero $\eta$. This obstructs reading a standard
finite asymptotic mass from that truncation, even though
$\ln r/r\to0$. We do not assign it a physical first law or a global
scalar QNM spectrum. There is no winding $\Tr'/\Tr$ to compute in
this branch, since $\Tr$ is identically zero.

\section{An exact branch with nonzero matter trace}
\subsection{Original matter action and its off-shell trace}

To activate the trace sector, choose the polynomial model
\begin{equation}
 \mathcal L_m(\mathcal S)=-\frac{\mathcal S+\zeta\mathcal S^2}{4\pi},
 \qquad
 \mathcal S=\frac14\mathcal F_{\mu\nu}\mathcal F^{\mu\nu},
 \qquad \mathcal F_{\mu\nu}=2\partial_{[\mu}A_{\nu]} .
 \label{NED}
\end{equation}
The invariant $\mathcal S$ must not be confused with an entropy.
We treat Eq.~\eqref{NED} as a specified classical polynomial theory,
not as the complete QED Euler--Heisenberg action. Its relation to
quadratic electromagnetic models is useful for comparison
\cite{Liang2026}. If regarded instead as a low-field truncation,
the additional condition $|\zeta\mathcal S|\ll1$ is required.

Let $K_{\mu\nu}=\mathcal F_{\mu\rho}\mathcal F_\nu{}^\rho$.
At fixed covariant $\mathcal F_{\mu\nu}$,
\begin{equation}
 \delta\mathcal S=\frac12K_{\mu\nu}\delta g^{\mu\nu},
 \qquad g^{\mu\nu}K_{\mu\nu}=4\mathcal S.
\end{equation}
Consequently,
\begin{align}
 T_{\mu\nu}&=-\mathcal L_{m,\mathcal S}K_{\mu\nu}
                    +\mathcal L_m g_{\mu\nu},\\
 \Tr&=4(\mathcal L_m-\mathcal S\mathcal L_{m,\mathcal S})
   =4\left[-\frac{\mathcal S+\zeta\mathcal S^2}{4\pi}
     +\frac{\mathcal S(1+2\zeta\mathcal S)}{4\pi}\right]
   =\frac{\zeta\mathcal S^2}{\pi},
 \label{trace-NED}\\
 D_{\mu\nu}&=\frac{\partial\Tr}{\partial\mathcal S}
              \frac{\partial\mathcal S}{\partial g^{\mu\nu}}
        =\frac{\zeta\mathcal S}{\pi}K_{\mu\nu}.
 \label{D-NED}
\end{align}
Equations~\eqref{trace-NED}--\eqref{D-NED} determine $\Theta=D-T$
without importing a perfect-fluid formula for $\Theta$.

Define
\begin{equation}
 C_\lambda=1-4\lambda,\qquad \zeta_\eff=\zeta C_\lambda.
 \label{C-definition}
\end{equation}
The off-shell identity
\begin{align}
 \mathcal L_m+\lambda\Tr
 &=-\frac{\mathcal S+\zeta\mathcal S^2}{4\pi}
       +\frac{\lambda\zeta\mathcal S^2}{\pi}\nonumber\\
 &=-\frac{\mathcal S+\zeta_\eff\mathcal S^2}{4\pi}
       \equiv\mathcal L_\eff
 \label{L-effective}
\end{align}
shows that the trace coupling renormalizes the quadratic matter
coefficient. Substitution of Eqs.~\eqref{trace-NED}--\eqref{D-NED}
into Eq.~\eqref{effective-tensor} independently gives
\begin{equation}
 T^\eff_{\mu\nu}
 =\frac{1+2\zeta_\eff\mathcal S}{4\pi}K_{\mu\nu}
   -\frac{\mathcal S+\zeta_\eff\mathcal S^2}{4\pi}g_{\mu\nu}.
 \label{T-effective-NED}
\end{equation}
Varying $A_\nu$, integrating by parts, and discarding the boundary
variation yields
\begin{equation}
 \LC_\mu\left(\mathcal L_{\eff,\mathcal S}
                   \mathcal F^{\mu\nu}\right)=0,
 \qquad \partial_{[\rho}\mathcal F_{\mu\nu]}=0.
 \label{gauge-equations}
\end{equation}
The effective matter equation, rather than the uncoupled original
matter equation, is the one required by the full action.

\subsection{Explicit metric and independent connection}

The coincident gauge must be specified in a coordinate system; setting
the connection to zero independently after changing coordinates is
not a covariant transformation \cite{Bahamonde2022}.
Use Cartesian coordinates $(\tau,x,y,z)$ and define
\begin{equation}
 r=(x^2+y^2+z^2)^{1/2},\quad n_i=x_i/r,\quad
 k_\mu\dd x^\mu=\dd\tau+\dd r,
 \quad k_\mu=(1,n_i).
\end{equation}
For $r>0$ take
\begin{equation}
 \Gamma^\rho{}_{\mu\nu}=0,\qquad
 g_{\mu\nu}=\eta_{\mu\nu}+H(r)k_\mu k_\nu,
 \qquad H=1-F.
 \label{KS}
\end{equation}
The one-form $k$ is null with respect to $\eta$ and is radially
geodesic. The matrix determinant lemma and nullness give
\begin{equation}
 \det g=-1,\qquad
 g^{\mu\nu}=\eta^{\mu\nu}-H k^\mu k^\nu.
 \label{KS-inverse}
\end{equation}
For completeness, the contractions that establish $Q=0$ are displayed
explicitly. By spherical symmetry evaluate at $(x,y,z)=(r,0,0)$,
retaining transverse derivatives. Then $k_\mu=(1,1,0,0)$ and the
nonzero $Q_{\alpha\mu\nu}=\partial_\alpha g_{\mu\nu}$ are
\begin{align}
 Q_{100}=Q_{101}=Q_{110}=Q_{111}&=H',\\
 Q_{202}=Q_{220}=Q_{212}=Q_{221}&=H/r,\\
 Q_{303}=Q_{330}=Q_{313}=Q_{331}&=H/r.
 \label{KS-Qcomponents}
\end{align}
The $(0,1)$ block of the inverse metric is
\begin{equation}
 (g^{ab})_{a,b=0,1}=
 \begin{pmatrix}-1-H&H\\H&1-H\end{pmatrix},
 \qquad g^{22}=g^{33}=1.
\end{equation}
Contraction of Eq.~\eqref{KS-Qcomponents} gives
\begin{align}
 Q_\alpha&=0,&
 \widetilde Q_\alpha&=(H'+2H/r)k_\alpha,\\
 Q_{\alpha\mu\nu}Q^{\alpha\mu\nu}&=0,&
 Q_{\alpha\mu\nu}Q^{\mu\alpha\nu}&=0.
 \label{KS-contractions}
\end{align}
For example, $Q_1=H'(g^{00}+2g^{01}+g^{11})=0$;
the same null contractions cancel the quadratic terms. Thus
Eq.~\eqref{Q-definition} gives
\begin{equation}
 Q=0\quad\hbox{for arbitrary differentiable }H(r).
 \label{Q-zero}
\end{equation}
This is vanishing of the scalar $Q$, not vanishing of the nonmetricity
tensor. Equations~\eqref{model-field} and \eqref{connection} now reduce
exactly to
\begin{equation}
 \mathring G_{\mu\nu}=8\pi T^\eff_{\mu\nu},
 \qquad \hbox{the linear-$Q$ connection identity}.
 \label{Einstein-branch}
\end{equation}
In particular, every $\alpha Q^2$ contribution to the background first
variation is zero. This does not prove that $\alpha$ drops out of
coupled perturbations: the second variation can contain
$\alpha(\delta Q)^2$.

The Cartesian Kerr--Schild metric is equivalently
\begin{equation}
 \dd s^2=-F\dd\tau^2+2(1-F)\dd\tau\dd r
                +(2-F)\dd r^2+r^2\dd\Omega^2.
\end{equation}
In a static exterior introduce
\begin{equation}
 \dd\tau=\dd t+\frac{1-F}{F}\dd r.
\end{equation}
The cross term then cancels and the radial coefficient is
$(1-F)^2/F+2-F=1/F$, giving
\begin{equation}
 \dd s^2=-F(r)\dd t^2+\frac{\dd r^2}{F(r)}+r^2\dd\Omega^2.
 \label{static-metric}
\end{equation}
The connection in these static spherical coordinates is the coordinate
transform of the connection in Eq.~\eqref{KS}, rather than a newly
imposed zero connection. If $X^A=(\tau,x,y,z)$ are the above affine
coordinates, it is explicitly defined by
\begin{equation}
 \Gamma^\rho{}_{\mu\nu}
 =\frac{\partial x^\rho}{\partial X^A}
   \frac{\partial^2 X^A}{\partial x^\mu\partial x^\nu}.
 \label{transformed-connection}
\end{equation}
This formula supplies the independent connection throughout every
static coordinate patch where the transformation is invertible.

\subsection{Magnetic solution of the matter and metric equations}

Take the magnetic two-form
\begin{equation}
 \mathcal F_{\theta\phi}=q\sin\theta,
 \qquad q=\frac{1}{4\pi}\int_{S^2}\mathcal F.
 \label{magnetic}
\end{equation}
It is represented by the usual two local monopole potentials and is
closed for $r>0$. Raising the indices using Eq.~\eqref{static-metric},
\begin{equation}
 \mathcal F^{\theta\phi}=\frac{q}{r^4\sin\theta},
 \qquad \mathcal S=\frac{q^2}{2r^4}.
 \label{magnetic-invariant}
\end{equation}
For $\nu=\phi$, the only nontrivial derivative in the first of
Eqs.~\eqref{gauge-equations} is proportional to
\begin{equation}
 \partial_\theta
  \left[r^2\sin\theta\,\mathcal L_{\eff,\mathcal S}
        \frac{q}{r^4\sin\theta}\right]
 =\partial_\theta
        \left[\frac{q\mathcal L_{\eff,\mathcal S}}{r^2}\right]=0.
\end{equation}
The other components vanish or reduce to a $\phi$ derivative of a
$\phi$-independent quantity. Thus the gauge equation is solved.

Because $K^t{}_t=K^r{}_r=0$ and
$K^\theta{}_\theta=K^\phi{}_\phi=q^2/r^4=2\mathcal S$,
Eq.~\eqref{T-effective-NED} gives
\begin{align}
 T^{\eff\,\mu}{}_{\nu}
     &=\operatorname{diag}(-\rho,-\rho,p_\perp,p_\perp),\\
 \rho&=\frac{q^2}{8\pi r^4}
       +\frac{\zeta_\eff q^4}{16\pi r^8},&
 p_\perp&=\frac{q^2}{8\pi r^4}
       +\frac{3\zeta_\eff q^4}{16\pi r^8}.
 \label{density-pressure}
\end{align}
The relation $T^{\eff\,t}{}_t=T^{\eff\,r}{}_r$ is also consistent
with unit lapse: in
$\dd s^2=-e^{2\delta}N\dd t^2+N^{-1}\dd r^2+r^2\dd\Omega^2$,
the difference of the two Einstein equations gives
$2N\delta'/r=8\pi(T^{\eff\,r}{}_r-T^{\eff\,t}{}_t)=0$.
The asymptotic normalization fixes the remaining constant to zero.

Set $F=1-2m(r)/r$. Direct differentiation gives
\begin{equation}
 \mathring G^t{}_t=\mathring G^r{}_r
 =\frac{rF'+F-1}{r^2}=-\frac{2m'}{r^2}.
\end{equation}
Using Eq.~\eqref{density-pressure}, the $tt$ equation integrates as
\begin{align}
 m'&=4\pi r^2\rho
       =\frac{q^2}{2r^2}+\frac{\zeta_\eff q^4}{4r^6},\\
 m(r)&=M-\frac{q^2}{2r}-\frac{\zeta_\eff q^4}{20r^5},\\
 F(r)&=1-\frac{2M}{r}+\frac{q^2}{r^2}+\frac{d}{r^6},
 \qquad d=\frac{\zeta_\eff q^4}{10}.
 \label{exact-F}
\end{align}
The angular equation is independently satisfied:
\begin{equation}
 \mathring G^\theta{}_\theta
 =\frac{F''}{2}+\frac{F'}r
 =\frac{q^2}{r^4}+\frac{15d}{r^8}
 =8\pi p_\perp.
 \label{angular-check}
\end{equation}
All mixed components vanish. Moreover,
$\rho'+2(\rho+p_\perp)/r=0$, as required by
$\LC_\mu T^{\eff\,\mu}{}_r=0$.
The original and effective traces are respectively
\begin{equation}
 \Tr(r)=\frac{\zeta q^4}{4\pi r^8},\qquad
 \Tr_\eff=C_\lambda\Tr.
 \label{trace-profile}
\end{equation}
The Ricci scalar provides an additional independent check:
\begin{align}
 \mathring R&=-F''-\frac{4F'}r-\frac{2(F-1)}{r^2}
             =-\frac{20d}{r^8},\\
 \mathring R&=-8\pi\Tr_\eff
             =-\frac{2\zeta_\eff q^4}{r^8}.
 \label{Ricci-check}
\end{align}
These expressions agree upon substituting $d$.
Thus Eq.~\eqref{exact-F}, together with
Eqs.~\eqref{KS}, \eqref{magnetic}, and
\eqref{transformed-connection}, is an exact background of the
specified polynomial theory for $r>0$.
It has a central singularity and is not a regular-core construction.

\subsection{Admissible region and parameter degeneracy}

We select parameters for which $F$ has a largest positive simple root
$r_h$, $F>0$ outside it, and $F'(r_h)>0$. We use the Killing time
normalized by $F\to1$ at infinity. For $\zeta_\eff\geq0$, the
displayed effective density is positive; this observation alone does
not establish stability or causal propagation of every matter and
gravity mode. No such general stability claim is needed below.

The metric depends on $\zeta$ and $\lambda$ only through
$\zeta_\eff=\zeta(1-4\lambda)$. Hence metric observables and the
massless scalar frequencies considered here cannot determine those
two coefficients separately. At $\lambda=1/4$,
\begin{equation}
 \mathcal L_\eff=-\frac{\mathcal S}{4\pi},\qquad d=0,
 \qquad F=1-\frac{2M}{r}+\frac{q^2}{r^2},
 \label{cancellation}
\end{equation}
even when the original matter trace in Eq.~\eqref{trace-profile} is
nonzero. This is exact cancellation of its nonlinear contribution in
the total action, not the disappearance of the original trace.
It is one useful test of a proposed winding-based interpretation.

\section{Hawking temperature and horizon residues}
\subsection{Euclidean regularity with a general radial lapse}

For clarity, start with
$\dd s^2=-A(r)\dd t^2+\dd r^2/B(r)+r^2\dd\Omega^2$.
Assume a common simple horizon with $A'_h,B'_h>0$.
Writing $x=r-r_h$ and $x=B'_h\varrho^2/4$, the leading Euclidean
two-metric is
\begin{equation}
 \dd s_E^2=A'_h x\dd\tau_E^2+\frac{\dd x^2}{B'_h x}
 =\dd\varrho^2+\frac{A'_hB'_h}{4}\varrho^2\dd\tau_E^2.
\end{equation}
Regularity at the origin requires
\begin{equation}
 \beta_H=\frac{4\pi}{\sqrt{A'_hB'_h}},\qquad
 T_H=\frac{\sqrt{A'_hB'_h}}{4\pi}.
 \label{general-temperature}
\end{equation}
The optical coordinate satisfies
\begin{equation}
 r_* =\int\frac{\dd r}{\sqrt{AB}},\qquad
 \frac1{\sqrt{AB}}=
 \frac{1}{\sqrt{A'_hB'_h}}\frac1{x}+O(1).
\end{equation}
The residue formula therefore refers to $1/\sqrt{AB}$ in a
two-function metric. The exact branch above has $A=B=F$, giving
\begin{equation}
 \mathcal R_h=\Res_{z=r_h}\frac1{F(z)}=\frac1{F'(r_h)},
 \qquad T_H=\frac{1}{4\pi\mathcal R_h}.
 \label{residue-identity}
\end{equation}
The square-root branch is chosen by the real static exterior.
Under a nonsingular analytic change of radial coordinate, it is the
residue of the differential $\dd r/F$ that is invariant. A rescaling
of Killing time rescales the temperature, so the time normalization
must always be stated.

\subsection{Euler-characteristic check and the boundary term}

The RVB construction uses the Euler characteristic
\cite{Robson2019}. On the nonextremal Euclidean disk truncated at
$r=R$, $\sqrt{g_2}=1$ and $R_2=-F''$.
The outward boundary has
$k=F'(R)/(2\sqrt{F(R)})$ and
$\dd s=\sqrt{F(R)}\dd\tau_E$. Gauss--Bonnet gives
\begin{align}
 \chi&=\frac1{4\pi}\int R_2\sqrt{g_2}\dd^2x
            +\frac1{2\pi}\int_{r=R}k\dd s\nonumber\\
 &=\frac{\beta_H}{4\pi}[F'(r_h)-F'(R)]
       +\frac{\beta_H}{4\pi}F'(R)
 =\frac{\beta_H F'(r_h)}{4\pi}=1.
\end{align}
Thus the same $\beta_H$ follows when the boundary is included.
This is a statement for one smooth nonextremal disk. Multiple horizons
with unequal temperatures need not admit one globally regular period.
The trace winding introduced below is a different index from $\chi$.

\subsection{Explicit temperature of the magnetic branch}

The horizon condition and the derivatives of Eq.~\eqref{exact-F} are
\begin{align}
 2M&=r_h+\frac{q^2}{r_h}+\frac{d}{r_h^5},
 \label{horizon-M}\\
 F'&=\frac{2M}{r^2}-\frac{2q^2}{r^3}-\frac{6d}{r^7},&
 F''&=-\frac{4M}{r^3}+\frac{6q^2}{r^4}+\frac{42d}{r^8}.
 \label{F-derivatives}
\end{align}
Substituting Eq.~\eqref{horizon-M} into $F'(r_h)$ gives
\begin{align}
 F'(r_h)&=\frac1{r_h}
          \left(1-\frac{q^2}{r_h^2}-\frac{5d}{r_h^6}\right),\\
 \mathcal R_h&=\frac{r_h^7}{r_h^6-q^2r_h^4-5d},\\
 T_H&=\frac{1}{4\pi r_h}
          \left(1-\frac{q^2}{r_h^2}-\frac{5d}{r_h^6}\right).
 \label{temperature-exact}
\end{align}
The denominator in $\mathcal R_h$ is nonzero on the assumed simple
branch. Its zero indicates degeneracy and invalidates the derivation
based on a simple pole.
The limits $\lambda\to0$, $\zeta\to0$, and $q\to0$ recover,
respectively, the polynomial Einstein--NED geometry, the
Reissner--Nordstrom geometry, and Schwarzschild.
The cancellation limit in Eq.~\eqref{cancellation} is a separate,
nonzero-bare-trace route to Reissner--Nordstrom.

\subsection{Controlled expansion about a nonextremal RN solution}

At fixed $M,q$, with $M>|q|$, define
\begin{equation}
 r_\pm=M\pm\sqrt{M^2-q^2},\qquad
 \Delta=r_+-r_->0,\qquad
 F_0=1-\frac{2M}{r}+\frac{q^2}{r^2}.
\end{equation}
For small $d$ write $r_h=r_++\delta r_h+O(d^2)$, where the
$O(d^2)$ notation holds at fixed dimensional background scales.
Taylor expansion gives
\begin{equation}
 0=F_0'(r_+)\delta r_h+\frac{d}{r_+^6}+O(d^2),
 \qquad F_0'(r_+)=\frac{\Delta}{r_+^2},
\end{equation}
and therefore
\begin{equation}
 \delta r_h=-\frac{d}{r_+^4\Delta}.
 \label{horizon-shift}
\end{equation}
Next,
\begin{align}
 F'(r_h)&=F_0'(r_+)+F_0''(r_+)\delta r_h
                             -\frac{6d}{r_+^7}+O(d^2),\\
 F_0''(r_+)&=\frac{-2r_++4r_-}{r_+^3},\\
 T_H&=T_{RN}-\frac{d(2r_+-r_-)}{2\pi r_+^7\Delta}+O(d^2),
 \qquad T_{RN}=\frac{\Delta}{4\pi r_+^2},
 \label{temperature-shift}\\
 \frac{T_H-T_{RN}}{T_{RN}}
 &=-\frac{2d(2r_+-r_-)}{r_+^5\Delta^2}+O(d^2).
 \label{temperature-relative}
\end{align}
Both $|d|/(r_+^5\Delta)\ll1$ and
$|d|(2r_+-r_-)/(r_+^5\Delta^2)\ll1$ must hold for the horizon and
relative-temperature expansions. The second condition is especially
important near extremality; the expansion is not uniform as
$\Delta\to0$.

\section{Complex algebra of the matter-trace sector}
\subsection{Divisors, logarithmic derivatives and winding}

An analytic continuation of a real solution is an additional local
regularity assumption. It holds explicitly for the rational trace in
Eq.~\eqref{trace-profile}. For a non-identically-zero meromorphic
$\Tr(z)$, the Laurent and residue theorems give
\cite{DLMF}
\begin{equation}
 \Tr(z)=(z-a)^m u(z),\qquad u(a)\ne0,
 \qquad \frac{\Tr'}{\Tr}=\frac{m}{z-a}+\frac{u'}u,
 \label{factor-local}
\end{equation}
where $m=\ord_a\Tr$ is positive for a zero and negative for a pole.
The nonvanishing factor $u$ carries the remaining continuous
information. Thus
\begin{equation}
 \Res_a\left(\frac{\Tr'}{\Tr}\dd z\right)=m.
 \label{trace-order}
\end{equation}
For a positively oriented simple contour $\gamma$ with no zero or
pole on it, the argument principle reads
\begin{equation}
 \nu_\Tr(\gamma)=\frac{1}{2\pi\ii}
             \oint_\gamma\frac{\Tr'(z)}{\Tr(z)}\dd z
       =N_{\rm zero}-N_{\rm pole}\in\mathbb Z.
 \label{winding}
\end{equation}
The count includes multiplicities. It is the winding of the image
curve $\Tr(\gamma)$ around zero. Equivalently it is the monodromy of
$\log(\Tr/\Tr_{\rm ref})/(2\pi\ii)$, where $\Tr_{\rm ref}$ is a
fixed nonzero quantity of the same dimension as $\Tr$.
The trace itself remains single-valued when meromorphic.

On a simply connected neighborhood with finitely many zeros and poles,
the full representation after factoring the divisor is
\begin{equation}
 \Tr(z)=\Tr_{\rm ref}\,e^{b(z)}
   \frac{\prod_j[(z-a_j)/\ell_0]^{m_j}}
        {\prod_k[(z-p_k)/\ell_0]^{n_k}},
 \label{trace-factorization}
\end{equation}
where $b$ is holomorphic and $\ell_0$ is a reference length.
The integer orders do not determine $b$, the scale, or the matter
variation $D_{\mu\nu}$. For example,
$\nu_{c\Tr}=\nu_\Tr$ for every nonzero constant $c$.
Nor does an essential singularity admit the finite divisor description
in Eq.~\eqref{factor-local}. These restrictions are part of the
domain of the winding description.

\subsection{Exact winding of the magnetic trace}

For $\zeta q^4\ne0$ the analytic continuation is
\begin{equation}
 \Tr(z)=\frac{A_\Tr}{z^8},\qquad
 A_\Tr=\frac{\zeta q^4}{4\pi},\qquad
 \frac{\Tr'}{\Tr}=-\frac8z.
 \label{trace-meromorphic}
\end{equation}
Consequently,
\begin{equation}
 \nu_\Tr(\gamma)=-8\,\operatorname{Ind}(\gamma,0).
 \label{trace-winding-exact}
\end{equation}
A once-counterclockwise contour enclosing the central pole has
$\nu_\Tr=-8$. This integer characterizes the pole order, not the
magnetic flux $q$ and not a quantization of the trace amplitude.
The contour around zero is a complex-analytic diagnostic; it is not
a physical loop through the singular center.

If a small contour $\gamma_h$ encloses only a horizon $r_h>0$, then
$\Tr$ is finite, holomorphic and nonzero inside it, so
\begin{equation}
 \nu_\Tr(\gamma_h)=0,
 \qquad
 \nu_F(\gamma_h)=\frac{1}{2\pi\ii}\oint_{\gamma_h}
                                     \frac{F'}F\dd z=1.
 \label{different-contours}
\end{equation}
These are different maps on the same contour. The matter-trace pole
at the origin must not be treated as if it were a horizon pole.
When $\zeta=0$ or $q=0$, $\Tr\equiv0$ and $\Tr'/\Tr$ is
undefined; its winding is not assigned the value zero by substitution.
Likewise, $\Tr_\eff\equiv0$ at $\lambda=1/4$, although the bare
trace still has winding $-8$ for $\zeta q^4\ne0$.

\begin{proposition}[Winding does not determine temperature]
Within the nonextremal magnetic family at fixed nonzero $q,\zeta$,
the trace profile and its winding remain unchanged as $\lambda$
varies, while the horizon residue and temperature generally change.
\end{proposition}
\begin{proof}
Equation~\eqref{trace-meromorphic} is independent of $\lambda$, whereas
$d=\zeta(1-4\lambda)q^4/10$ enters
Eqs.~\eqref{horizon-M} and \eqref{temperature-exact}.
For a nonextremal RN seed, the leading variation in
Eq.~\eqref{temperature-shift} is nonzero because
$2r_+-r_->0$ and $\partial_\lambda d=-2\zeta q^4/5\ne0$.
Thus identical trace winding occurs with different temperatures.
The numerical example below also evaluates the exact roots directly.
\end{proof}

This example gives a precise role to the proposed algebraic structure:
the trace sector can be represented by its divisor together with a
nonvanishing analytic factor. The field equations still require the
full trace, the action, its metric response, and the coupling constants.

\subsection{Why horizon multiplicity must accompany a residue}

Even the geometric relation in Eq.~\eqref{residue-identity} has a
necessary multiplicity hypothesis. If
$F=a_2(z-r_h)^2+a_3(z-r_h)^3+\cdots$, then
\begin{equation}
 \frac1F=\frac{1}{a_2(z-r_h)^2}
       -\frac{a_3}{a_2^2(z-r_h)}+O(1),
 \qquad \nu_F(\gamma_h)=2.
\end{equation}
The residue can be finite even though the surface gravity vanishes.
An explicit counterexample is extremal Reissner--Nordstrom:
\begin{equation}
 F=\left(1-\frac Mr\right)^2,\qquad
 \frac1F=\frac{M^2}{(r-M)^2}+\frac{2M}{r-M}+1.
\end{equation}
Here $\Res_M(1/F)=2M$ but $T_H=F'(M)/(4\pi)=0$.
Using $1/(4\pi\Res(1/F))$ would give a spurious nonzero
temperature. The order of the horizon zero is therefore indispensable.
The integer $\nu_F$ and the length-valued simple-horizon residue
$\mathcal R_h$ encode different information; neither is an arbitrary
extra term to be added to the temperature.

\section{Scalar propagation and local monodromy}
\subsection{A probe consistent with the trace coupling}

To specify which field is perturbed, introduce a real massless scalar
of small amplitude $\varepsilon$ in the original matter action:
\begin{equation}
 \mathcal L_m^{\rm total}
 =\mathcal L_m^{\rm NED}-\frac{\varepsilon^2}{2}X,
 \qquad X=g^{\mu\nu}\partial_\mu\Phi\partial_\nu\Phi.
\end{equation}
Its original stress tensor and trace are
\begin{equation}
 T^{(\Phi)}_{\mu\nu}=\varepsilon^2
 (\partial_\mu\Phi\partial_\nu\Phi-\tfrac12g_{\mu\nu}X),
 \qquad \Tr^{(\Phi)}=-\varepsilon^2 X.
\end{equation}
The quadratic scalar action in the \emph{full} theory is therefore
\begin{equation}
 I^{(2)}_\Phi=-\frac{\varepsilon^2}{2}(1+2\lambda)
      \int\sqrt{-g}\,X\dd^4x.
 \label{scalar-action}
\end{equation}
For $1+2\lambda>0$, a constant field normalization gives
\begin{equation}
 \mathring\Box\Phi
 =\frac{1}{\sqrt{-g}}\partial_\mu
    (\sqrt{-g}g^{\mu\nu}\partial_\nu\Phi)=0.
 \label{KG}
\end{equation}
At $\lambda=-1/2$ this quadratic equation is degenerate; that point is
excluded. Negative kinetic sign is also excluded for this probe.
The scalar changes the background only at $O(\varepsilon^2)$.
Thus holding the background fixed while calculating the linear
scalar wave is consistent. The trace coupling does not add a
separate radial kinetic factor in this linear-$\Tr$ model.

\subsection{Every step to the radial master equation}

For Eq.~\eqref{static-metric}, $\sqrt{-g}=r^2\sin\theta$.
Substitution into Eq.~\eqref{KG} gives
\begin{equation}
 -\frac1F\partial_t^2\Phi
 +\frac1{r^2}\partial_r(r^2F\partial_r\Phi)
 +\frac1{r^2}\Delta_{S^2}\Phi=0.
\end{equation}
Separate
\begin{equation}
 \Phi=e^{-\ii\omega t}Y_{\ell m}(\theta,\phi)\frac{\psi(r)}r,
 \qquad \Delta_{S^2}Y_{\ell m}=-\ell(\ell+1)Y_{\ell m}.
\end{equation}
The radial differentiation is
\begin{align}
 \partial_r(\psi/r)&=\psi'/r-\psi/r^2,\\
 r^2F\partial_r(\psi/r)&=F(r\psi'-\psi),\\
 \partial_r[F(r\psi'-\psi)]&=F'(r\psi'-\psi)+Fr\psi''.
\end{align}
After multiplication by $Fr$ the equation becomes
\begin{equation}
 F^2\psi''+FF'\psi'
 +\left\{\omega^2-F\left[\frac{\ell(\ell+1)}{r^2}
                                  +\frac{F'}r\right]\right\}\psi=0.
\end{equation}
Define $\dd r_*/\dd r=1/F$. Since
$\partial_{r_*}^2\psi=F^2\psi''+FF'\psi'$, the master equation is
\begin{equation}
 \frac{\dd^2\psi}{\dd r_*^2}
       +[\omega^2-V_\ell(r)]\psi=0,
 \qquad
 V_\ell=F\left[\frac{\ell(\ell+1)}{r^2}+\frac{F'}r\right].
 \label{master}
\end{equation}
For the exact magnetic branch,
\begin{equation}
 V_\ell=\left(1-\frac{2M}r+\frac{q^2}{r^2}+\frac d{r^6}\right)
 \left[\frac{\ell(\ell+1)}{r^2}
    +\frac{2M}{r^3}-\frac{2q^2}{r^4}-\frac{6d}{r^8}\right].
 \label{explicit-potential}
\end{equation}
This is a scalar potential, not a gravitational or electromagnetic
master potential.

\subsection{Boundary conditions and the horizon continuation}

At a simple outer horizon,
\begin{equation}
 F=F'_h(r-r_h)+O((r-r_h)^2),\qquad
 r_* =\mathcal R_h\log[(r-r_h)/\ell_0]+O(1).
\end{equation}
The scalar potential vanishes there and the two local solutions are
$\psi_\pm\sim e^{\pm\ii\omega r_*}$.
With the time convention $e^{-\ii\omega t}$ the ingoing solution is
\begin{equation}
 \psi_{\rm in}\sim e^{-\ii\omega r_*}
     \sim[(r-r_h)/\ell_0]^{-\ii\omega\mathcal R_h}.
\end{equation}
A counterclockwise complex loop around the horizon changes
$\log(r-r_h)$ by $2\pi\ii$, so its radial monodromy is
\begin{equation}
 \mathcal M_{\rm in}=e^{2\pi\omega\mathcal R_h},\qquad
 T_H=\frac1{4\pi\mathcal R_h}.
 \label{monodromy}
\end{equation}
This radial continuation, with $t$ held fixed, is a local analytic
identity; it is not by itself an emission probability or a global
frequency quantization condition.
For the asymptotically flat exact branch, the QNM boundary conditions
are ingoing at $r_*\to-\infty$ and outgoing
$\psi\sim e^{+\ii\omega r_*}$ at $r_*\to+\infty$.
They select global discrete frequencies only after the radial equation
is solved throughout the exterior. The horizon index
$\nu_\Tr(\gamma_h)=0$ introduces no additional local wave monodromy.

\section{Metric null orbit and scalar eikonal shifts}
\subsection{Leading WKB and the domain of the eikonal approximation}

For a single smooth potential barrier the leading WKB approximation
is \cite{Schutz1985}
\begin{equation}
 \omega^2\simeq V_0-\ii(n+\tfrac12)\sqrt{-2V_0''},
 \label{WKB}
\end{equation}
where the subscript $0$ denotes the barrier maximum and the double
prime here means $\dd^2/\dd r_*^2$. At the maximum,
$V_0''=F(r_0)^2\dd^2V/\dd r^2|_{r_0}$.
We use only the scalar eikonal limit $\ell\gg1$ at fixed $n$.
Then $V\simeq L^2F/r^2$ with $L=\ell+1/2$, and its leading maximum
satisfies
\begin{equation}
 \left(\frac{F}{r^2}\right)'_{r_c}=0,
 \qquad r_cF'(r_c)-2F(r_c)=0.
 \label{orbit-equation}
\end{equation}
For null geodesics of the metric,
$\dot r^2+J^2F/r^2=E^2$. Setting this radial potential and its first
derivative to their circular-orbit values gives
\begin{equation}
 \Omega_c=\frac{\sqrt{F_c}}{r_c},\qquad
 \Lambda_c^2=
 \frac{F_c[2F_c-r_c^2F_c'']}{2r_c^2}.
 \label{orbit-frequencies}
\end{equation}
For the second formula, expand the radial equation to second order
in a displacement and convert affine time to coordinate time using
$\dot t=E/F_c$ and $E^2=J^2F_c/r_c^2$.
These equations yield the leading scalar correspondence
\cite{Cardoso2009}
\begin{equation}
 \omega_{\ell n}=L\Omega_c-\ii(n+\tfrac12)\Lambda_c+O(L^{-1}).
 \label{eikonal}
\end{equation}
An unstable outer orbit and the single-barrier regime are assumed.
In nonlinear electrodynamics the characteristics of electromagnetic
waves can involve an effective metric \cite{Liang2026}. Our orbit is
the \emph{metric} null orbit appropriate to this scalar eikonal limit;
we do not identify it with a complete electromagnetic shadow calculation.

\subsection{Exact orbit equations in the magnetic branch}

Substitution of Eqs.~\eqref{exact-F} and \eqref{F-derivatives} into
Eq.~\eqref{orbit-equation} gives, term by term,
\begin{equation}
 -2+\frac{6M}{r_c}-\frac{4q^2}{r_c^2}-\frac{8d}{r_c^6}=0,
 \quad\Longleftrightarrow\quad
 r_c^6-3Mr_c^5+2q^2r_c^4+4d=0.
 \label{orbit-polynomial}
\end{equation}
The same derivative substitution gives
\begin{equation}
 2F-r^2F''=2-\frac{4q^2}{r^2}-\frac{40d}{r^6},
\end{equation}
so that
\begin{equation}
 \Omega_c=\frac{\sqrt{F_c}}{r_c},\qquad
 \Lambda_c=\Omega_c
      \sqrt{1-\frac{2q^2}{r_c^2}-\frac{20d}{r_c^6}}.
 \label{exact-eikonal-factors}
\end{equation}
These are exact orbit diagnostics for the specified metric.
Equation~\eqref{eikonal}, in contrast, remains an approximation to
the scalar spectrum.

\subsection{First-order shifts at fixed mass and magnetic charge}

Let the outer RN orbit and its associated factors be
\begin{equation}
 r_0=\frac{3M+\sqrt{9M^2-8q^2}}2,\quad
 D_c=2r_0-3M,\quad
 F_{0c}=\frac{r_0^2-q^2}{3r_0^2},\quad
 K_0=1-\frac{2q^2}{r_0^2}.
\end{equation}
Divide Eq.~\eqref{orbit-polynomial} by $r_c^4$ and expand around
$r_c=r_0+\delta r_c$:
\begin{equation}
 r_c^2-3Mr_c+2q^2+\frac{4d}{r_c^4}=0
 \quad\Longrightarrow\quad
 D_c\delta r_c+\frac{4d}{r_0^4}=O(d^2).
\end{equation}
Thus
\begin{equation}
 \delta r_c=-\frac{4d}{r_0^4D_c}.
 \label{orbit-shift}
\end{equation}
Because $(F_0/r^2)'_{r_0}=0$, the orbit displacement does not
contribute to the first variation of $\Omega_c^2$:
\begin{align}
 \delta(\Omega_c^2)
 &=\left(\frac{F_0}{r^2}\right)'_{r_0}\delta r_c
       +\frac{d}{r_0^8}=\frac{d}{r_0^8},\\
 \frac{\delta\Omega_c}{\Omega_0}
 &=\frac{d}{2r_0^6F_{0c}}
   =\frac{3d}{2r_0^4(r_0^2-q^2)}.
 \label{omega-shift}
\end{align}
For $\Lambda_c=\Omega_c\sqrt K$,
\begin{equation}
 \delta K=\frac{4q^2}{r_0^3}\delta r_c-\frac{20d}{r_0^6}.
\end{equation}
Hence, with $\Lambda_0=\Omega_0\sqrt{K_0}$,
\begin{align}
 \frac{\delta\Lambda_c}{\Lambda_0}
 &=\frac{\delta\Omega_c}{\Omega_0}+\frac{\delta K}{2K_0}\nonumber\\
 &=\frac{d}{2r_0^6F_{0c}}
   -\frac{10d}{K_0r_0^6}
   -\frac{8q^2d}{K_0r_0^7D_c}.
 \label{lambda-shift}
\end{align}
Equations~\eqref{temperature-relative}, \eqref{omega-shift}, and
\eqref{lambda-shift} sample different radial data. A common horizon
residue does not replace the exterior-potential calculation.
The expansion requires $|\delta r_c|\ll r_0$ and small relative
frequency corrections, as well as the nondegenerate-orbit condition
$D_c\ne0$ and positive $K_0$.

\section{Reproduction and limits of the claims}

 Table~\ref{numerical-table} uses exact numerical roots of
Eqs.~\eqref{exact-F} and \eqref{orbit-polynomial}; the entries are
local thermal and orbit quantities, not finite-$\ell$ QNM eigenvalues.
All parameters are expressed in units of $M$, with
$q/M=0.6$ and $\zeta/M^2=1$. The bare trace winding around the origin
is $-8$ in every row.

\begin{table}[htbp]
\centering
\small
\caption{Exact background diagnostics at $q/M=0.6$, $\zeta/M^2=1$.
The quantities $M T_H$, $M\Omega_c$, and $M\Lambda_c$ are dimensionless.
The last row is the exact cancellation limit $d=0$.}
\label{numerical-table}
\begin{tabular}{@{}rrrrrr@{}}
\toprule
$\lambda$ & $r_h/M$ & $MT_H$ & $r_c/M$ & $M\Omega_c$ & $M\Lambda_c$\\
\midrule
0.00 & 1.79922669 & 0.03922570 & 2.73655798 & 0.20582662 & 0.19561529\\
0.10 & 1.79953643 & 0.03925448 & 2.73670752 & 0.20582262 & 0.19563936\\
0.20 & 1.79984561 & 0.03928319 & 2.73685699 & 0.20581862 & 0.19566342\\
0.25 & 1.80000000 & 0.03929752 & 2.73693169 & 0.20581662 & 0.19567544\\
\bottomrule
\end{tabular}
\end{table}

For an independent small-$d$ numerical check, subtract the analytic
linear term from each exact root or observable at
$d/M^6=10^{-3}$ and $5\times10^{-4}$. The remainder ratios are
$4.00054$ for $r_h$, $4.00061$ for $T_H$, $4.00017$ for $r_c$,
$4.00014$ for $\Omega_c$, and $4.00021$ for $\Lambda_c$.
Their proximity to four checks the predicted quadratic remainder.
Midpoint integration with $16384$ points separately reproduces
$\nu_\Tr=-8$ on a contour enclosing zero,
$\nu_\Tr=0$ on a small horizon contour, and
$\nu_F=1$ on that horizon contour.

Several boundaries of the result are essential. First, the magnetic
branch is $Q=0$, so it does not measure the nonlinear nonmetricity
coupling $\alpha$. The vacuum logarithmic branch has a different
connection and cannot be added to it. Second, the linear trace coupling
in the selected NED model is exactly absorbed into $\zeta_\eff$;
separate identification of $\lambda$ needs additional physical input.
Third, winding is a property of a specified analytically continued
function and contour. It is neither a new matter degree of freedom
nor a general reconstruction of an energy-momentum tensor.
Fourth, the coupled metric--connection--electromagnetic perturbation
system has not been solved. Scalar eikonal diagnostics cannot establish
its stability or gravitational spectrum. Finally, no physical entropy,
heat capacity, or thermodynamic geometry is inferred by integrating a
metric-parameter equation. Such claims require an appropriate charge
and variational analysis \cite{Wald1993}.

\section{Conclusion}

For the specified theory $f(Q,\Tr)=Q+\alpha Q^2+16\pi\lambda\Tr$,
the vacuum logarithmic branch retains the local residue--temperature
relation of quadratic $f(Q)$ gravity. A nonzero matter trace requires
an explicit matter sector. Polynomial nonlinear electrodynamics gives
an exact magnetic branch with
\begin{equation}
 Q=0,\qquad \Tr=\frac{\zeta q^4}{4\pi r^8},\qquad
 F=1-\frac{2M}{r}+\frac{q^2}{r^2}
       +\frac{\zeta(1-4\lambda)q^4}{10r^6}.
\end{equation}
All independent background field equations and the connection
realization have been checked. The simple-horizon residue determines
the Euclidean temperature and the local massless-scalar monodromy;
the exterior metric determines the distinct scalar eikonal shifts.

The proposed winding description becomes precise through
$\dd\log\Tr$: it supplies the integer divisor of the meromorphic
trace, while a nonvanishing analytic factor retains the missing
continuous information. In the exact example, the central trace
winding is $-8$ for continuously different temperatures, and its
local horizon winding is zero. Horizon multiplicity must also be
checked before using an inverse-residue temperature formula.
These explicit distinctions give a controlled algebraic description
of the trace sector without identifying $\Tr$ with a winding number.
Extending the result to a branch with both $Q\ne0$ and $\Tr\ne0$, or
to genuinely mixed $Q\Tr$ couplings, requires a new coupled solution.

\appendix
\section{Direct variational route for the quadratic nonmetricity term}

The Einstein form of the metric equation can also be recovered without
memorizing a general $f(Q)$ equation. For the chosen convention,
write the linear-$Q$ variation as
\begin{equation}
 \delta(\sqrt{-g}Q)
 =\sqrt{-g}\,\mathring G_{\mu\nu}\delta g^{\mu\nu}
       +\partial_\alpha B^\alpha,
 \qquad
 B^\alpha=-2\sqrt{-g}P^\alpha{}_{\mu\nu}\delta g^{\mu\nu},
\end{equation}
up to equivalent boundary representatives for the same fixed-connection
variation. Since
\begin{equation}
 \delta(\sqrt{-g}Q^2)
 =2Q\,\delta(\sqrt{-g}Q)-Q^2\delta\sqrt{-g},
\end{equation}
the bulk terms before the second integration by parts are
\begin{equation}
 \sqrt{-g}\left(2Q\mathring G_{\mu\nu}
             +\tfrac12Q^2g_{\mu\nu}\right)\delta g^{\mu\nu}
       +2Q\partial_\alpha B^\alpha.
\end{equation}
Now use
$2Q\partial_\alpha B^\alpha=
 \partial_\alpha(2QB^\alpha)-2(\partial_\alpha Q)B^\alpha$.
Discarding the total derivative gives
\begin{equation}
 \frac{\delta\int\sqrt{-g}Q^2\dd^4x}{\sqrt{-g}\,\delta g^{\mu\nu}}
 =2Q\mathring G_{\mu\nu}+\frac12Q^2g_{\mu\nu}
                         +4P^\alpha{}_{\mu\nu}\partial_\alpha Q.
\end{equation}
Adding the linear-$Q$ and trace variations reproduces
Eqs.~\eqref{model-field}--\eqref{effective-tensor}.
This also shows directly why $Q=0$ eliminates the background
quadratic contribution while allowing a nonzero second variation.

\section{Information carried by the three contour quantities}

Table~\ref{contour-table} records the maps and hypotheses separately.
The trace in the last two rows is the original matter trace.

\begin{table}[htbp]
\centering
\small
\caption{Distinct complex-analytic data in the exact branch.}
\label{contour-table}
\begin{tabular}{@{}>{\raggedright\arraybackslash}p{0.28\textwidth}>{\raggedright\arraybackslash}p{0.20\textwidth}>{\raggedright\arraybackslash}p{0.42\textwidth}@{}}
\toprule
Quantity & Value & Information and domain\\
\midrule
$\Res_{r_h}(\dd z/F)$ & $1/F'_h$ & A length-valued coefficient; inverse
temperature only for a simple horizon and specified Killing time.\\[3pt]
$\nu_F(\gamma_h)$ & $1$ & Order of a simple horizon zero; becomes $2$
at a double zero.\\[3pt]
$\nu_\Tr(\gamma_h)$ & $0$ & Trace regular and nonzero near a positive
horizon.\\[3pt]
$\nu_\Tr(\gamma_0)$ & $-8$ & Eighth-order central pole for
$\zeta q^4\ne0$ and one positive circuit about zero.\\
\bottomrule
\end{tabular}
\end{table}


\begin{thebibliography}{99}

\bibitem{Jimenez2018}
J. Beltr\'an Jim\'enez, L. Heisenberg, and T. Koivisto,
``Coincident general relativity,''
Phys. Rev. D \textbf{98}, 044048 (2018).
\href{https://doi.org/10.1103/PhysRevD.98.044048}{doi:10.1103/PhysRevD.98.044048}.
\href{https://arxiv.org/abs/1710.03116}{arXiv:1710.03116}.

\bibitem{DAmbrosio2022}
F. D'Ambrosio, S. D. B. Fell, L. Heisenberg, and S. Kuhn,
``Black holes in $f(\mathbb Q)$ gravity,''
Phys. Rev. D \textbf{105}, 024042 (2022).
\href{https://doi.org/10.1103/PhysRevD.105.024042}{doi:10.1103/PhysRevD.105.024042}.
\href{https://arxiv.org/abs/2109.03174}{arXiv:2109.03174}.

\bibitem{Xu2019}
Y. Xu, G. Li, T. Harko, and S.-D. Liang,
``$f(Q,T)$ gravity,'' Eur. Phys. J. C \textbf{79}, 708 (2019).
\href{https://doi.org/10.1140/epjc/s10052-019-7207-4}{doi:10.1140/epjc/s10052-019-7207-4}.
\href{https://arxiv.org/abs/1908.04760}{arXiv:1908.04760}.

\bibitem{Chen2026}
W.-X. Chen, ``Horizon residues and test-scalar quasinormal modes of a
perturbative Schwarzschild-type black hole in quadratic $f(Q)$ gravity,''
Eur. Phys. J. C \textbf{86}, 991 (2026).
\href{https://doi.org/10.1140/epjc/s10052-026-16260-4}{doi:10.1140/epjc/s10052-026-16260-4}.

\bibitem{Hawking1975}
S. W. Hawking, ``Particle creation by black holes,''
Commun. Math. Phys. \textbf{43}, 199--220 (1975).
\href{https://doi.org/10.1007/BF02345020}{doi:10.1007/BF02345020}.

\bibitem{Robson2019}
C. W. Robson, L. Di Mauro Villari, and F. Biancalana,
``Topological nature of the Hawking temperature of black holes,''
Phys. Rev. D \textbf{99}, 044042 (2019).
\href{https://doi.org/10.1103/PhysRevD.99.044042}{doi:10.1103/PhysRevD.99.044042}.
\href{https://arxiv.org/abs/1810.09322}{arXiv:1810.09322}.

\bibitem{Chen2025}
W.-X. Chen, ``Calculating the Hawking temperature of black holes in
$f(Q)$ gravity using the RVB method: a residue-based approach,''
Can. J. Phys. \textbf{103}, 531--542 (2025).
\href{https://doi.org/10.1139/cjp-2024-0214}{doi:10.1139/cjp-2024-0214}.
\href{https://arxiv.org/abs/2501.17485}{arXiv:2501.17485}.

\bibitem{Bronnikov2001}
K. A. Bronnikov, ``Regular magnetic black holes and monopoles from
nonlinear electrodynamics,'' Phys. Rev. D \textbf{63}, 044005 (2001).
\href{https://doi.org/10.1103/PhysRevD.63.044005}{doi:10.1103/PhysRevD.63.044005}.
\href{https://arxiv.org/abs/gr-qc/0006014}{arXiv:gr-qc/0006014}.

\bibitem{Harko2011}
T. Harko, F. S. N. Lobo, S. Nojiri, and S. D. Odintsov,
``$f(R,T)$ gravity,'' Phys. Rev. D \textbf{84}, 024020 (2011).
\href{https://doi.org/10.1103/PhysRevD.84.024020}{doi:10.1103/PhysRevD.84.024020}.
\href{https://arxiv.org/abs/1104.2669}{arXiv:1104.2669}.

\bibitem{Liang2026}
Y. Liang, J. Tao, and R. Yang,
``Black holes in $f(R,T)$ gravity coupled with Euler--Heisenberg
electrodynamics,'' Phys. Lett. B \textbf{872}, 140092 (2026).
\href{https://doi.org/10.1016/j.physletb.2025.140092}{doi:10.1016/j.physletb.2025.140092}.
\href{https://arxiv.org/abs/2506.14860}{arXiv:2506.14860}.

\bibitem{Bahamonde2022}
S. Bahamonde and L. J\"arv,
``Coincident gauge for static spherical field configurations in symmetric
teleparallel gravity,'' Eur. Phys. J. C \textbf{82}, 963 (2022).
\href{https://doi.org/10.1140/epjc/s10052-022-10922-9}{doi:10.1140/epjc/s10052-022-10922-9}.
\href{https://arxiv.org/abs/2208.01872}{arXiv:2208.01872}.

\bibitem{DLMF}
NIST Digital Library of Mathematical Functions,
\S1.10, ``Functions of a Complex Variable,'' especially
\S1.10(iii)--(iv), Laurent series and the residue theorem.
\url{https://dlmf.nist.gov/1.10} (accessed 16 October 2026).

\bibitem{Schutz1985}
B. F. Schutz and C. M. Will,
``Black hole normal modes: A semianalytic approach,''
Astrophys. J. Lett. \textbf{291}, L33--L36 (1985).
\href{https://doi.org/10.1086/184453}{doi:10.1086/184453}.

\bibitem{Cardoso2009}
V. Cardoso, A. S. Miranda, E. Berti, H. Witek, and V. T. Zanchin,
``Geodesic stability, Lyapunov exponents and quasinormal modes,''
Phys. Rev. D \textbf{79}, 064016 (2009).
\href{https://doi.org/10.1103/PhysRevD.79.064016}{doi:10.1103/PhysRevD.79.064016}.
\href{https://arxiv.org/abs/0812.1806}{arXiv:0812.1806}.

\bibitem{Wald1993}
R. M. Wald, ``Black hole entropy is the Noether charge,''
Phys. Rev. D \textbf{48}, R3427--R3431 (1993).
\href{https://doi.org/10.1103/PhysRevD.48.R3427}{doi:10.1103/PhysRevD.48.R3427}.
\href{https://arxiv.org/abs/gr-qc/9307038}{arXiv:gr-qc/9307038}.

\end{thebibliography}
\end{document}